\documentclass[twocolumn,showpacs,preprintnumbers,amsmath,amssymb]{revtex4}
\usepackage{graphicx}% Include figure files
\usepackage{dcolumn}% Align table columns on decimal point
\usepackage{bm}% bold math

\newcommand{\ket}[1]{|#1\rangle}
\newcommand{\bra}[1]{\langle#1|}

\begin{document}

\title{Total correlations and mutual information}
\author{Zbigniew Walczak}
 \email{Z.Walczak@merlin.phys.uni.lodz.pl}
\affiliation{%
Department of Theoretical Physics, University of Lodz\\
Pomorska 149/153, 90-236 {\L}\'od\'z, Poland}

\begin{abstract}
In quantum information theory it is generally accepted
that quantum mutual information is an information-theoretic 
measure of total correlations of a bipartite quantum state. 
We argue that there exist quantum states for which quantum
mutual information cannot be considered as a measure of total 
correlations. Moreover, for these states we propose a different way 
of quantifying total correlations.
\end{abstract}
 

\maketitle

\section{Introduction}
In quantum information theory it is generally accepted that
(i) total correlations of a bipartite quantum state 
are quantified by quantum mutual information  
(see e.g. \cite{Stratonovich66, Zurek83, Barnett89, Barnett91,
Cerf97, Cerf98, Henderson01, Vedral02, Ollivier02,  Vedral03, 
Oppenheim03, Hamieh03, Hamieh04, Groisman05, Horodecki05,
Schumacher06, Li07, Chan07, Luo08, Wolf08, Usha08}) 
\begin{equation}
\label{eq:1}
{\cal T}(\rho^{AB}) = I(\rho^{AB}), 
\end{equation}
(ii) the total correlations ${\cal T}(\rho^{AB})$ can be decomposed
into classical and quantum correlations (see e.g. \cite{Cerf97,
Cerf98, Henderson01, Vedral02, Ollivier02,  Vedral03, Oppenheim03,
Hamieh03, Hamieh04, Groisman05, Horodecki05, Li07, Chan07, Luo08}) 
\begin{equation}
\label{eq:2}
{\cal T}(\rho^{AB}) = {\cal C}(\rho^{AB}) + {\cal Q}(\rho^{AB}),
\end{equation} 
(iii) the quantum correlations are dominated by the classical
correlations 
(see e.g. \cite{Henderson01, Vedral02, Vedral03, Hamieh03, 
Hamieh04, Groisman05, Horodecki05, Li07}) 
\begin{equation}
\label{eq:3}
{\cal C}(\rho^{AB}) \geq {\cal Q}(\rho^{AB}).
\end{equation} 
Notice that the quantum correlations are not only
limited to quantum entanglement,
because separable quantum states can also have
correlations 
which are responsible for the improvements of some quantum tasks that
cannot be simulated by classical methods 
\cite{Braunstein99, Meyer00, Biham04, Knill98, Datta05, Datta07,
Bennett99}.

However, there are some results that may raise doubts as to
whether the statements (i), (ii) and (iii) hold for all $\rho^{AB}$. 
In the following we briefly review and discuss a few of them. 
For example, it has been shown that there are 
quantum states for which statements (i) and (ii) cannot be
simultaneously true, if quantum correlations are measured by 
relative entropy of entanglement while classical correlations are
quantified by a measure based on the maximum information that could 
be extracted on one system by making a POVM measurement on the other 
one \cite{Henderson01}.

Recently, it has been shown that for certain quantum states
the quantum correlations, as measured by entanglement of formation, 
exceed half of the total correlations, as measured by quantum mutual 
information, 
${\cal Q}(\rho^{AB}) \geq \tfrac{1}{2} I(\rho^{AB})$ \cite{Li07}. 
If one assumes that statements (i), (ii) and (iii) hold for all 
$\rho^{AB}$, then one concludes that entanglement of formation cannot 
be considered as a measure of quantum correlations. However, if one
assumes that entanglement of formation is a measure of quantum
correlations and statements (ii) and (iii) hold, then it can be shown
that for these quantum states 
${\cal T}(\rho^{AB}) \geq I(\rho^{AB})$.

Moreover, it has been shown that for certain quantum states
quantum correlations, as measured by entanglement of formation,
exceed total correlations, as measured by quantum mutual
information, ${\cal Q}(\rho^{AB}) \geq {\cal T}(\rho^{AB})$ 
\cite{Hayden06, Li07}. It means that entanglement of formation 
cannot be considered as a measure of quantum correlations, 
if one assumes that statements (i), (ii) and
(iii) hold for all $\rho^{AB}$. 
However, if one assumes that entanglement of formation is a measure of
quantum correlations and statements (ii) and (iii) hold, then again
one comes to conclusion that for these quantum states 
${\cal T}(\rho^{AB}) \geq I(\rho^{AB})$.

Furthermore, it has been shown that for some quantum states 
quantum correlations, as measured by entanglement of formation, exceed
$I(\rho^{AB})- {\cal Q}(\rho^{AB})$, i.e. 
${\cal Q}(\rho^{AB}) \geq \tfrac{1}{2} I(\rho^{AB})$ \cite{Chan07}. 
If one assumes that statements (i) and (ii) hold for all $\rho^{AB}$,
then one concludes that for these states     
${\cal Q}(\rho^{AB}) \geq {\cal C}(\rho^{AB})$.
However, if one assumes that entanglement of formation is 
a measure of quantum
correlations and statements (ii) and (iii) hold, 
then once again one comes to conclusion that for these quantum states 
${\cal T}(\rho^{AB}) \geq I(\rho^{AB})$.

These results suggest that statement (i) may not be true for all
bipartite quantum states. It is clear that if (\ref{eq:1}) holds for all
states, then one immediately concludes that quantum
mutual information must be a measure of total correlations 
also in the case when the quantum state has only classical 
correlations. This implies that
\begin{equation}
\label{eq:CM}
{\cal C}(\rho^{AB}) = {\cal T}(\rho^{AB}) = I(\rho^{AB})
\end{equation}
for classically correlated quantum states.

In this Letter, we argue that there exist classically 
correlated quantum states for which quantum mutual information cannot 
be considered as a proper measure of total correlations and for these 
states we propose a different way of quantifying total correlations.

\section{A two-qubit mixed state}
Assume that Alice and Bob share a pair of qubits in the following
state  
\begin{equation}
\rho^{AB} = \alpha \ket{00}\bra{00} +
(1 - \alpha) \ket{11} \bra{11},
\label{stan qubitow}
\end{equation}
where $\alpha \in (0, 1)$. This state cannot have quantum correlations
because it is separable and qubit $A$ ($B$) is in the state 
$\rho^{A(B)}=\alpha \ket{0}\bra{0} + (1 - \alpha) \ket{1} \bra{1}$ 
which is a mixture of orthogonal states $\ket{0}$ and $\ket{1}$. 
Therefore, it is clear that the correlations between two orthogonal
states of qubits $A$ and $B$ can be purely classical 
(see e.g. \cite{Ollivier02, Luo08, Datta08}).   

Suppose now that Alice and Bob measure two observables  
$M_{A} = a_{0} \ket{0} \bra{0} + a_{1} \ket{1} \bra{1}$ 
and 
$M_{B} = b_{0} \ket{0} \bra{0} + b_{1} \ket{1} \bra{1}$, respectively.
If the measurement outcome of $M_{A}$ ($M_{B}$) 
is $a_{i}$ ($b_{i}$), then qubit $A$ ($B$) is certainly in the 
state $\ket{i}$.    
Therefore, it is clear that the classical correlations between two
orthogonal states of qubits $A$ and $B$ are simply correlations
between two classical random variables $A$ and $B$ corresponding to 
the measurement outcomes of $M_{A}$ and $M_{B}$, respectively.
Notice that $A$ and $B$ are random variables with alphabets 
${\cal A} = \{a_0, a_1 \}$, ${\cal B} = \{b_0, b_1 \}$
and probability mass functions $p^{A(B)} = [ p^{A (B)}_{i} ]$, where
$p^{A (B)}_{i} = \text{Tr} [\ket{i}\bra{i} \rho^{A (B)}]$ denotes 
the probability that the measurement outcome of $M_{A}$ 
($M_{B}$) is $a_{i}$ ($b_{i}$). In the case under consideration, it
can be shown that 
\begin{subequations}
\begin{eqnarray}
&& p^{A} = (\alpha, 1-\alpha), \\
&& p^{B} = (\alpha, 1-\alpha). 
\end{eqnarray}
\end{subequations}

Assume now that first Alice performs a measurement of $M_{A}$ 
and then Bob performs a measurement of $M_{B}$.
If the measurement outcome of $M_{A}$ is $a_{i}$,
then the post-measurement state of the system is given by 
$\rho^{AB}_{i} = (\ket{i}\bra{i} \otimes I) \rho^{AB} 
(\ket{i}\bra{i} \otimes I)/p^{A}_{i}$.
Therefore, the conditional probability that Bob's outcome is
$b_{j}$ provided that Alice's was $a_{i}$ is 
$p^{B|A}_{j|i} = \text{Tr}[(I \otimes \ket{j}\bra{j}) \rho^{AB}_{i}]$ 
while the joint probability that the measurement outcome of
$M_{A}$ and $M_{B}$ is $a_{i}$ and $b_{j}$, respectively 
is given by
$p^{AB}_{ij} = 
%p^{B|A}_{j|i} p^{A}_{i} =
\text{Tr}[(\ket{i}\bra{i} \otimes \ket{j}\bra{j}) \rho^{AB}]$.
In the case under consideration, it can be shown that 
\begin{equation}
\label{prawdopodobienstwa warunkowe dla qubitow}
p^{B|A} = [ p^{B|A}_{j|i}] = 
\left(
\begin{array}{cc}
1 & 0 \\
0 & 1
\end{array}
\right)
\end{equation}
and  
\begin{equation}
p^{AB} = [ p^{AB}_{ij}] = 
\left(
\begin{array}{cc}
\alpha & 0 \\
0 & 1-\alpha
\end{array}
\right).
\end{equation}
Thus, we see that random variables $A$ and $B$ are not independent  
and therefore there exist classical correlations between qubits 
$A$ and $B$ in the state (\ref{stan qubitow}). 

Assume now, according to the statement (i), 
that total correlations of a bipartite quantum state 
are quantified by quantum mutual information which is defined 
in formal analogy to classical 
mutual information as 
\begin{equation}
I(\rho^{AB}) = S(\rho^{A}) +S(\rho^{B}) - S(\rho^{AB}),
\end{equation} 
where $S(\cdot)$ denotes the von Neumann entropy.
Then, according to Eq.~(\ref{eq:CM}) 
the classical correlations between qubits $A$ and $B$  
are measured by the quantum mutual information. 
In the case under consideration, it can be shown that  
$I(\rho^{AB})$ is just classical mutual information of random 
variables $A$ and $B$ given by   
\begin{align}
I(A:B) & = \sum_{i,j}  p^{AB}_{ij} \log_{2}
(p^{AB}_{ij}/(p^{A}_{i} p^{B}_{j}))  \nonumber\\
& =  -\alpha \log_{2} \alpha - (1-\alpha) \log_{2} (1-\alpha).
\end{align}
Therefore, we see that the classical correlations content of the
quantum state (\ref{stan qubitow})  can be arbitrarily small, 
as measured by classical mutual information, 
because $\alpha \in (0,1)$.

\begin{figure}
  \centering
  \includegraphics[width=0.49\textwidth]{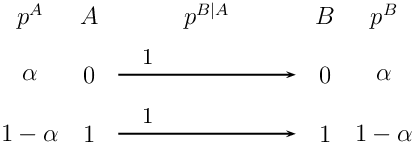}
   \caption{This diagram shows that $B = f(A)$ and therefore random 
   variables $A$ and $B$ are perfectly correlated in the 
   information-theoretic sense,
   despite the fact that in this case the mutual information
   can be arbitrarily small.}
  \label{fig1}
\end{figure}

Now, we check if these correlations can be really arbitrarily small.  
From Eq.~(\ref{prawdopodobienstwa warunkowe dla qubitow}) it follows
that if Alice's measurement outcome is $a_{i}$,
then Bob's one is $b_{i}$, i.e. $B$ is a one-to-one function of $A$, 
$B=f(A)$ (see Fig.~\ref{fig1}). 
It means that the random variables $A$ and $B$ and what follows the
states of qubits are perfectly correlated in the information-theoretic
sense.
Therefore, we see that in the general case quantum mutual information 
cannot be considered as a measure of total correlations in bipartite 
quantum states.  

Now, we show why the classical mutual information $I(A:B)$ does not
capture all the correlations between random variables $A$ and $B$, 
except the case when $\alpha = 1/2$.  
We know that Shannon entropy of random variable
$B$, $H(B) = - \sum_{j} p^{B}_{j} \log_{2} p^{B}_{j}$, is a measure
of Alice's {\em a priori} uncertainty about the measurement outcome of
$M_{B}$ and if the measurement outcome of $M_{A}$ is $a_{i}$, then
Alice's uncertainty about the measurement outcome of $M_B$ is changed,
preferably reduced,  to 
$H(B|A=a_{i}) = - \sum_{j} p^{B|A}_{j|i} \log_{2} p^{B|A}_{j|i}$. 
Therefore, the  information she gained about the measurement outcome
of $M_B$ due to the measurement of $M_{A}$ is given by 
$H(B) - H(B|A=a_{i})$.
Thus, the average information gain about the measurement outcome of
$M_{B}$ due to the knowledge of the measurement outcome of 
$M_{A}$ is $\sum_{i} p^{A}_{i} (H(B) - H(B|A=a_{i})) = H(B) - H(B|A)$,
and it can be shown that it is equal to $I(A:B)$. 
Therefore, we see that the average information gain about one
random variable due to the knowledge of other one can be
arbitrarily small although they are perfectly correlated.
Thus it is clear that classical mutual information is not a measure of
correlations between two random variables, it is rather a measure of
their mutual dependency. 
 
This conclusion leads us to the following question: What is an
information-theoretic measure of correlations between random variables
$A$ and $B$? 
For a pair of random variables with identical probability mass
functions Cover and Thomas \cite{Cover06} define it in the following
way 
\begin{equation}
{\cal C}(A,B) = \frac{I(A:B)}{H(A)}.
\end{equation}
The basic properties of ${\cal C}(A,B)$ are as follows
\cite{Cover06}  
(i) ${\cal C}(A,B) = {\cal C}(B,A)$, 
(ii) $ 0 \leq {\cal C}(A,B) \leq 1$, 
(iii) ${\cal C}(A,B) = 0$ if and only if $A$ and $B$ are independent, and 
(iv) ${\cal C}(A,B) = 1$ if and only if $A$ and $B$ are perfectly
correlated in the information-theoretic sense. 

In the case under consideration, $p^{A} = p^{B}$ and it can be shown
that $I(A:B) = H(A)$, therefore ${\cal C}(A,B) = 1$,
i.e. $A$ and $B$ are perfectly correlated for all 
$\alpha \in (0, 1)$. 
In the next section we show how to extend this
definition to the case when the probability mass functions are not
identical.

\section{A two-qutrit mixed state}
Assume now that Alice and Bob share a pair of qutrits in the following
separable state 
\begin{equation}
\rho^{AB} = \frac{1}{3} \ket{1 1}\bra{1 1} +
\frac{1}{3} \ket{2 0}\bra{2 0} +
\frac{1}{3} \ket{2 2}\bra{2 2}
\label{stanAB2}
\end{equation}
in which orthogonal states of qutrits $A$ and $B$ are classically
correlated. Notice that 
$\rho^{A} = \tfrac{1}{3} \ket{1}\bra{1} + \tfrac{2}{3} \ket{2}\bra{2}$
and
$\rho^{B} = \tfrac{1}{3} \ket{0}\bra{0} + \tfrac{1}{3} \ket{1}\bra{1}
+  \tfrac{1}{3} \ket{2}\bra{2}$.
Suppose now that Alice and Bob measure two observables  
$M_{A} = a_{0} \ket{0}\bra{0} + a_{1} \ket{1}\bra{1} + a_{2}
\ket{2}\bra{2}$ 
and  
$M_{B} = b_{0} \ket{0}\bra{0} + b_{1} \ket{1}\bra{1} + b_{2}
\ket{2}\bra{2}$.  
It can be easily shown that the probability mass functions 
$p^{A}$ and $p^{B}$ are not identical, and they are given by 
\begin{subequations}
\begin{eqnarray}
&& p^{A} = (0, \tfrac{1}{3}, \tfrac{2}{3}), \\
&& p^{B} = (\tfrac{1}{3}, \tfrac{1}{3}, \tfrac{1}{3}). 
\end{eqnarray}
\end{subequations}

\begin{figure}
  \centering
  \includegraphics[width=0.49\textwidth]{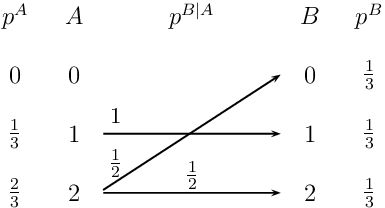}
   \caption{This diagram shows that the random variables $A$ and
    $B$ are not perfectly correlated in the information-theoretic
    sense.} 
  \label{fig2}
\end{figure}

Assume now that first Alice performs a measurement of $M_{A}$ and then
Bob performs a measurement of $M_{B}$.
It can be shown that
\begin{subequations}
\label{prawdopodobienstwa warunkowe dla qutritow}
\begin{eqnarray}
p^{B|A}_{0|1} = 0,  & \quad p^{B|A}_{1|1} = 1,  & \quad p^{B|A}_{2|1}
= 0, \\  
p^{B|A}_{0|2} = \tfrac{1}{2}, & \quad p^{B|A}_{1|2} = 0, & \quad 
p^{B|A}_{2|2} = \tfrac{1}{2},
\end{eqnarray}
\end{subequations}
and  
\begin{equation}
p^{AB} = [ p^{AB}_{ij}] = 
\left(
\begin{array}{ccc}
0 & 0 & 0 \\
0 & \frac{1}{3} & 0 \\
\frac{1}{3} & 0 & \frac{1}{3}
\end{array}
\right).
\label{prawdopodobienstwa laczne dla qutritow}
\end{equation} 
Thus, we see that the random variables $A$ and $B$ are not
independent, 
and from Eqs.~(\ref{prawdopodobienstwa warunkowe dla qutritow}) it 
follows that Bob's  measurement outcomes are correlated with Alice's 
ones, but they are not perfectly correlated (see Fig.~\ref{fig2}). 
Now, we show how to quantify these correlations. 
We know from classical information 
theory that for any two random variables $A$ and $B$ 
(i) $H(B|A) \leq H(B)$ with
equality if and only if they are independent, and
(ii) $H(B|A) \geq 0$ with
equality if and only if $B=f(A)$, i.e. 
$B$ is perfectly correlated 
with $A$ 
in the information-theoretic sense. 
Therefore, it is clear that if $H(B)>0$, then 
\begin{equation}
1 \geq H(B|A)/H(B) \geq 0.
\end{equation}
Notice that this inequality can be rewritten in the following form 
\begin{equation}
0 \leq I(A:B)/H(B) \leq 1. 
\end{equation}
Therefore, the correlations between $A$
and $B$ can be measured by $I(A:B)/H(B)$. In the case under
consideration, it can be shown that $I(A:B) = \log_{2} 3 -
\frac{2}{3}$, $H(B) = \log_{2} 3$ and therefore  
\begin{equation}
\frac{I(A:B)}{H(B)} = 1 - \tfrac{2}{3} (\log_{2} 3)^{-1}.
\end{equation}

\begin{figure}
  \centering
  \includegraphics[width=0.49\textwidth]{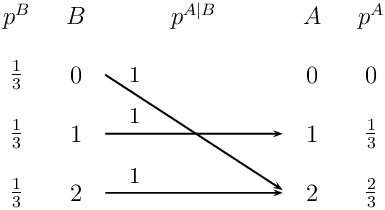}
   \caption{This diagram shows the perfect correlations, in the
   information-theoretic sense, between random variables $A$ and 
   $B$, $A=f(B)$, despite the fact that the mutual information 
   $I(A:B) = H(A) \simeq 0.918$.}
  \label{fig3}
\end{figure}

Assume now that first Bob performs a measurement of $M_{B}$ and then 
Alice performs a measurement of $M_{A}$. 
If the measurement outcome of
$M_{B}$ is $b_{i}$, then the post-measurement state of the system is 
given by  
$\rho^{AB}_{\phantom{A} i} = (I \otimes \ket{i}\bra{i}) \rho^{AB} 
(I \otimes \ket{i}\bra{i})/p^{B}_{i}$.
Therefore, the conditional probability that Alice's outcome is
$a_{j}$ provided that Bob's was $b_{i}$ is 
$p^{A|B}_{j|i} = \text{Tr}[(\ket{j}\bra{j} \otimes I) 
\rho^{AB}_{\phantom{A}i}]$
while the joint probability that the measurement outcome of
$M_{A}$ and $M_{B}$ is $a_{j}$ and $b_{i}$, respectively 
is given by 
$p^{AB}_{ji} = 
%p^{A|B}_{j|i} p^{B}_{i} =
\text{Tr}[(\ket{j}\bra{j} \otimes \ket{i}\bra{i}) \rho^{AB}]$.
It this case, the joint probabilities are given by
(\ref{prawdopodobienstwa laczne dla qutritow}) 
while the conditional probabilities are as follows  
\begin{equation}
p^{A|B} = [ p^{A|B}_{j|i}] = 
\left(
\begin{array}{ccc}
0 & 0 & 0 \\
0 & 1 & 0 \\
1 & 0 & 1
\end{array}
\right).
\end{equation}
Thus, we see that Alice's measurement outcomes are perfectly
correlated with Bob's ones (see Fig.~\ref{fig3}). 
Now, we explain why this is the case.
We know that for any two random variables $A$ and $B$
(i) $H(A|B) \leq H(A)$ with equality if and only if they are
independent, and  
(ii) $H(A|B) \geq 0$ with equality if and only if $A=f(B)$, 
i.e. $A$ 
is perfectly correlated with $B$. 
Therefore, it is clear that if $H(A)>0$, then 
\begin{equation}
1 \geq H(A|B)/H(A) \geq 0.
\end{equation}
Notice that this inequality can be rewritten in the following form 
\begin{equation}
0 \leq I(A:B)/H(A) \leq 1.
\end{equation} 
Therefore, the correlations between $A$
and $B$ can be measured by $I(A:B)/H(A)$. 
In the case under
consideration, it can be shown that $I(A:B) = H(A)$ and therefore 
\begin{equation}
\frac{I(A:B)}{H(A)} = 1.
\end{equation}

Thus, we see that the correlations between two random variables $A$
and $B$ corresponding to the measurement outcomes of $M_{A}$ and
$M_{B}$ depend on the temporal order of the measurements performed by
Alice and Bob.    
Therefore, in order to capture all classical correlations that can be
observed in the state (\ref{stanAB2}), we propose to define a measure
of correlations between two random variables $A$ and $B$ in the
following way 
\begin{align}
\label{eq:miara korelacji}
{\cal C}(A,B) & = 
{\rm Max}\left(\frac{I(A:B)}{H(A)},
\frac{I(A:B)}{H(B)}\right) = \nonumber \\
& = \frac{I(A:B)}{{\rm Min}(H(A),H(B))}.
\end{align}
It is easy to note that (\ref{eq:miara korelacji}) has
the following properties (i) ${\cal C}(A,B) = {\cal C}(B,A)$,   
(ii) $ 0 \leq {\cal C}(A,B) \leq 1$,
(iii) ${\cal C}(A,B) = 0$ if and only if $A$ and $B$ are independent, 
and (iv) ${\cal C}(A,B) = 1$ if and only if $A$ and $B$ are perfectly
correlated in the information-theoretic sense.
The first property follows directly from (\ref{eq:miara korelacji}). 
The second one follows from inequality 
$0 \leq I(A:B) \leq {\rm Min}(H(A),H(B))$.
The third property holds because $I(A:B) = 0$ if and only if $A$ and
$B$ are independent.
And the last property can be derived in the following way. 
If $A=f(B)$ then $H(A|B) = 0$ and $I(A:B) = H(A)$. 
Thus ${\cal C}(A,B) = H(A)/{\rm Min}(H(A),H(B))$ and taking into
account that $H(f(B)) \leq H(B)$ we get ${\cal C}(A,B) = 1$. In the
similar way one can show that if $B=f(A)$ then also ${\cal C}(A,B) = 1$. 
Conversely, assume that ${\cal C}(A,B) = 1$. 
Then $I(A:B) = {\rm Min}(H(A),H(B))$. 
If ${\rm Min}(H(A),H(B)) = H(A)$ then $H(A|B) = 0$ and this implies
that $A=f(B)$. 
And if ${\rm Min}(H(A),H(B)) = H(B)$ then $H(B|A) = 0$ and this implies
that $B=f(A)$. 
Moreover, it is easy to note that ${\cal C}(A,B)$ is equal
to $I(A:B)$ if and only if $I(A:B)=0$ or ${\rm Min}(H(A),H(B))=1$.

If a state $\rho^{AB}$ has only classical correlations
then the way of finding the correlations 
between two random variables $A$ and $B$ corresponding to 
the measurement outcomes of $M_{A}$ and $M_{B}$ can be considerably
simplified by noting that in this case the Shannon entropies are 
just equal to the von Neumann entropies and from 
Eqs.~(\ref{eq:CM}) and (\ref{eq:miara korelacji}) it follows that    
\begin{equation}
\label{eq:TC}
{\cal T}(\rho^{AB}) = 
\frac{I(\rho^{AB})}{{\rm Min}(S(\rho^{A}),S(\rho^{B}))}.
\end{equation}

Thus, we see that for the classically correlated quantum
states $\rho^{AB}$--in the general 
case it is difficult to see whether the state is classically
correlated 
or entangled one--total correlations should be
quantified by (\ref{eq:TC}) instead of quantum mutual information, 
except these cases for which 
${\rm Min}(S(\rho^{A}),S(\rho^{B}))=1$.

Now, it is natural to ask a fundamental question: 
Can we apply our results for entangled states?
Before we answer this question we should recall that 
beyond analyzing the correlations in the simplest examples
of states, the Bell-like states and the Bell states, still very little
is known in this complex subject.

In \cite{Barnett89} a measure of the total
correlations based on the 
correlation coefficient was introduced and it was shown that for the
Bell-like state 
$\ket{\psi^{+}} = \sqrt{\alpha}\ket{01} + 
\sqrt{1-\alpha}\, e^{i\varphi} \ket{10}$
this measure of the total correlations and quantum mutual information
have a single maximum at $\alpha=1/2$, and they are equal to zero 
when $\alpha=0$ or $\alpha=1$. Of course, this important result does
not mean that the quantum mutual information is a measure of the total
correlations in the Bell-like state. This suggests only a possible
connection between these two quantities. Moreover, we do not even know 
whether this property holds for other two-qubit states.

Recently, in \cite{Groisman05} it has been shown that for
the Bell state 
$\ket{\phi^{+}} = \frac{1}{\sqrt{2}}(\ket{00}+\ket{11})$ 
the quantum mutual information is equal to the amount of randomness
which is necessary to completely erase all the correlations in 
$\ket{\phi^{+}}$. This remarkable result gives us an operational
interpretation for the quantum mutual information, at least for the
Bell state. Moreover, it may shed new light on the connection between
the total correlations and the quantum mutual information.

Now, let us apply our results to the entangled states  
$\ket{\psi^{+}}$ and $\ket{\phi^{+}}$.
Assume now that  
(i) Alice and Bob share a pair of qubits in the state $\ket{\psi^{+}}$,  
(ii) Alice and Bob measure two observables  
$M_{A} = a_{0} \ket{0} \bra{0} + a_{1} \ket{1} \bra{1}$ 
and 
$M_{B} = b_{0} \ket{0} \bra{0} + b_{1} \ket{1} \bra{1}$, respectively,
and
(iii) first Alice performs a measurement of $M_{A}$ and then Bob
performs a measurement of $M_{B}$. 
In this case, it can be easily shown that (i) the probability mass
functions $p^{A}$ and $p^{B}$ are given by 
\begin{subequations}
\begin{eqnarray}
&& p^{A} = (\alpha, 1-\alpha), \\
&& p^{B} = (1-\alpha, \alpha), 
\end{eqnarray}
\end{subequations}
(ii) the conditional probabilities are as follows
\begin{equation}
\label{pw}
p^{B|A} = [ p^{B|A}_{j|i}] = 
\left(
\begin{array}{cc}
0 & 1 \\
1 & 0
\end{array}
\right),
\end{equation}
and (iii) the joint probabilities are given by  
\begin{equation}
p^{AB} = [ p^{AB}_{ij}] = 
\left(
\begin{array}{cc}
0 & \alpha \\
1-\alpha & 0
\end{array}
\right).
\end{equation}     
Thus, we see that Alice's measurement outcomes are perfectly
correlated with Bob's ones in the information-theoretic sense, which 
follows directly from (\ref{pw}). Moreover, it can be shown 
that this result will be the same if the sequence of measurements is
opposite. Therefore, the total correlations in the state 
$\ket{\psi^{+}}$, understood as correlations between the measurement 
outcomes of $M_{A}$ and $M_{B}$, are maximal for all $\alpha$ and
$\varphi$, not only for $\alpha = 1/2$ and all $\varphi$, 
as it was suggested in \cite{Barnett89}.

Notice that in the similar way it can be shown that
the total correlations in the state $\ket{\phi^{+}}$ are also
maximal. Of course, our result does not
contradict the result presented in \cite{Groisman05}, 
because in general for two maximally correlated quantum states
the amount of randomness needed to erase all the
correlations may be different. 
For example, it can be shown that for the maximally correlated two-qubit
state of the form (\ref{stan qubitow}) we need $1$ bit of randomness 
if $\alpha = 1/2$, and $2$ bits of randomness if 
$\alpha \neq 1/2$.

\section{Conclusion}
In this Letter, we have shown that for bipartite quantum systems 
there exist quantum states for which quantum mutual information 
cannot be considered as a universal measure 
of total correlations, understood as the correlations between 
the measurement outcomes of two local observables.  
Moreover, for these states we have proposed a different way of
quantifying total correlations, which takes into account that the 
correlations can depend on the temporal order of the measurements.         
Furthermore, we have shown that this way of
quantifying total correlations can be successfully applied  in the
case of the simplest examples of entangled states, which were
studied in the literature from the other points of view. 
However, it remains an open question whether the same can be done for
other non-classically correlated bipartite quantum states.

\begin{acknowledgments}
This work was supported by the University of Lodz Grant and by 
LFPPI network. 
\end{acknowledgments}

\bibliographystyle{unsrt}
%\bibliography{bibliography}

\end{document}